\documentclass{epl}

\usepackage{bm}
\usepackage{cite}
\usepackage{amsmath}
\usepackage{graphicx}

\title{Correlation length of hydrophobic polyelectrolyte solutions}
\shorttitle{Correlation length of hydrophobic PE}
\author{D. Baigl\inst{1} \and R. Ober\inst{1} \and D. Qu\inst{2}\thanks{Present address: Laboratoire de Physique des Solides, Universit\'e Paris Sud, Orsay, France} \and A. Fery\inst{2} \and C. E. Williams\inst{1}\thanks{E-mail: \email{claudine.williams@college-de-france.fr}}}
\shortauthor{D. Baigl \etal} \institute{
  \inst{1} Laboratoire des Fluides Organis\'es,
CNRS UMR 7125, Coll\`ege de France - 11, place Marcelin Berthelot,
75005 Paris, France.
 \\
 \inst{2} Max-Planck-Institut f\"{u}r Kolloid und
 Grenzfl\"{a}chenforschung,
Am M\"{u}hlenberg 1, 14476 Golm, Germany.}


\begin{document}

\maketitle

\begin{abstract}
The combination of two techniques (Small Angle X-ray Scattering
and Atomic Force Microscopy) has allowed us to measure in
reciprocal and real space the correlation length $\xi$ of
salt-free aqueous solutions of highly charged hydrophobic
polyelectrolyte as a function of the polymer concentration $C_p$,
charge fraction $f$ and chain length $N$. Contrary to the
classical behaviour of hydrophilic polyelectrolytes in the strong
coupling limit, $\xi$ is strongly dependent on $f$. In particular
a continuous transition has been observed from $\xi \sim
C_p^{-1/2}$ to $\xi\sim C_p^{-1/3}$ when $f$ decreased from 100\%
to 35\%. We interpret this unusual behaviour as the consequence of
the two features characterising the hydrophobic polyelectrolytes:
the pearl necklace conformation of the chains and the anomalously
strong reduction of the effective charge fraction.
\end{abstract}

\section{Introduction} Polyelectrolytes are polymers containing ionisable groups
which, in a polar solvent like water, dissociate into charges tied
to the polymer backbone and counter-ions spread in the solution.
Polyelectrolytes are called hydrophobic when water is a poor
solvent for the backbone. Their single chain properties
(conformation and effective charge density) are very different
from those of hydrophilic polyelectrolytes.
Theories~\cite{Kantor,Dobrynin_1996, Dobrynin_1999},
simulations~\cite{Micka_1999,Limbach_2002} and
experiments~\cite{CW_houches,Baigl_epl} are all consistent with
 a pearl-necklace description of the single chain where low
 dielectric constant nanoregions (the pearls) are connected by narrow
 elongated strings. Another important feature
 is the charge renormalisation. By osmometry Essafi \emph{et al.} measured
  the effective charge fraction $f_{eff}$
as a function of the chemical charge fraction $f$ for partially
sulfonated polystyrene (PSS)~\cite{Essafi_thesis}.
 In a range of charge fractions where hydrophilic polyelectrolytes followed rather
well the so-called Manning-Oosawa condensation theory
\cite{Manning_1969,Oosawa_1971}, they found for PSS a further
strong reduction of $f_{eff}$~\cite{feff_note}.

What happens for a large number of interacting chains in solution
is still a largely open question. Indeed in the framework of the
theoretical pearl-necklace model two semi-dilute regimes have been
predicted~\cite{Dobrynin_1999}. As long as the pearl size is much
smaller than the correlation length $\xi$ which characterises the
transient network of overlapping chains, a classical
polyelectrolyte behaviour is expected, \emph{i.e.} $\xi\sim
C_p^{-1/2}$ where $C_p$ is the polymer concentration
(string-controlled regime). However when the pearl size becomes of
the order of $\xi$, a transition to $\xi\sim C_p^{-1/3}$ is
predicted (bead-controlled regime). Experimentally two regimes
have only been observed for an atypical polyelectrolyte soluble in
polar solvents other than water, the scaling exponents being -1/2
and -1/7 respectively~\cite{Waigh_2001}. For the model hydrophobic
polyelectrolyte PSS in water, Essafi \emph{et al.} found by Small
Angle Scattering measurements (SAXS and SANS) that the position of
the so called polyelectrolyte peak, $q^*$, which is a measure of
the inverse of $\xi$, scaled as $q^*\sim C_{p}^{\alpha}$ in the
whole $C_p$ range with ${\alpha}$ depending qualitatively on
$f$~\cite{Essafi_thesis,Essafi_1993,Essafi_1995}. This behaviour
was later observed for more complex systems like concentrated
polyelectrolyte stars~\cite{Rawiso_2001}. On the other hand, using
thin film balance technique, Th\'eodoly \emph{et al.} measured the
wavelength $\lambda$ of the oscillating disjoining pressure
isotherms of various polyelectrolytes confined in thin liquid
films~\cite{Theodoly_langmuir}. They found for hydrophobic PSS the
same concentration dependence as for hydrophilic polyelectrolytes:
$\lambda \sim C_p^{-1/2}$.

In such an unclear situation it seemed essential to measure
systematically the correlation length as a function of $f$, chain
length $N$ and $C_p$. For this purpose well defined model
hydrophobic polyelectrolytes, PSS's, have been
prepared~\cite{Baigl_2002} and $\xi$ has been measured by two
different techniques: SAXS for a classical measurement in
reciprocal space and Atomic Force Microscopy (AFM) in real space.
In the SAXS experiment $\xi$ has been taken as $2\pi/q^*$ where
$q^*$ is the position of the polyelectrolyte peak. AFM has been
used to provide force-distance measurements of the polyelectrolyte
solution confined in a sphere-plane geometry. The period $d$ of
this oscillating force has been assumed to be a direct measurement
of $\xi$.

\section{Experimental details}Well
defined monodisperse ($M_W/M_N<1.3$) PSS's
(poly(styrene-\emph{co}-styrene sulfonate, sodium salt) of various
chemical charge fractions $f$ and chain lengths $N$
(Table~\ref{tablePSS}) have been prepared and characterised
according to a method described elsewhere~\cite{Baigl_2002}.
Salt-free water was produced by a milli-Q system (Millipore co.)
feeded with triply-distilled water. PSS solutions were prepared by
weight and left under stirring for a few days in airtight plastic
bottles before use.

\begin{table}
\caption{PSS samples of various chain lengths $N$, chemical charge
fractions $f$ and effective charge fractions $f_{eff}$. $f_{eff}$
is calculated from $f$ according to~\cite{feff_note}}
\label{tablePSS}
\begin{center}
\begin{largetabular}{lllcccrrr}
    $N$   &$f(\%)$ &$f_{eff}(\%)$ &$N$ &$f(\%)$ &$f_{eff}(\%)$ &$N$ &$f(\%)$ &$f_{eff}(\%)$
\\  120     &34     &3.4      &1320   &36 &4.3     &2520   &37 &4.8
\\  \cline{1-3}
    410     &39     &5.8      &1320   &53 &12      &2520   &54 &13
\\  410     &56     &14       &1320   &71 &21      &2520   &89 &30
\\  \cline{7-9}
    410     &71     &21       &1320   &91 &31      &5800   &57 &14
\\  410     &91     &31       &       &   &        &5800   &90 &30

\end{largetabular}
\end{center}
\end{table}

For the SAXS experiments cylindrical capillaries of 2 mm internal
diameter were filled with the PSS solutions and sealed by Teflon
tape. Measurements were performed at the European Synchrotron
Radiation Facility (ESRF, Grenoble, France) on BM26 beamline. We
used the highly monochromatic
($\Delta\lambda/\lambda\approx2\times10^{-4}$) X-ray beam at
$\lambda = 0.124$ nm (10.0 keV). The detector was a 1D quadrant
gas-filled detector situated at 7.8 m from the sample. The
scattering vectors $q$ ($q=4\pi/\lambda \sin(\theta/2)$, $\theta$
is the scattering angle) ranged from 0.1 to 0.9 $\rm{nm}^{-1}$.

Force-Distance measurements were carried out using a commercial
Atomic Force Microscope (Molecular Force Probe, Asylum Research,
USA) equipped with an inductive sensor (LVDT sensor). We used the
colloidal probe technique~\cite{Ducker_1991} in which the usual
sharp AFM tip was replaced by a colloidal silica particle of 7
$\rm \mu m$ diameter(Bangs Laboratories Inc.) that was glued to a
tipless cantilever ($\rm \mu Mash, CsC12 $). Clean silicon wafers
(Wacker, Germany) were used as substrates. Particles glued to the
cantilevers were cleaned directly before the measurement by plasma
cleaning for 20 minutes (Plasma cleaner PDC-32G, Harrick
Scientific, USA). Then a drop of the PSS solution was put onto the
substrate and the probing head (the silica particle) was also
immersed in the solution. Force-distance curves were obtained at a
pulling frequency of 0.2 Hz, corresponding to piezo speed from 200
nm/s  to 800 nm/s, depending on the range of the pulling distance.
Variation of the pulling speed in this range showed no change in
force-distance data. For each solution, the force-distance curves
were measured at 3-5 different locations on the same substrate as
well as on 2-4 different substrates.

\section{Results and discussion}
\begin{figure}
\onefigure[scale=0.5]{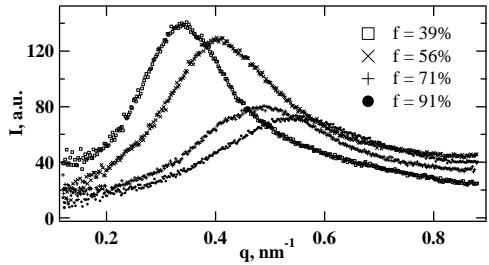} \caption{SAXS Intensity(arbitrary
units) as a function of the wavevector $q$ for PSS at $C_{p}=0.1$
mol/L, $N=410$ monomers and various chemical charge fractions
$f$.} \label{SAXS}
\end{figure}

\begin{figure}
\onefigure[scale=0.5]{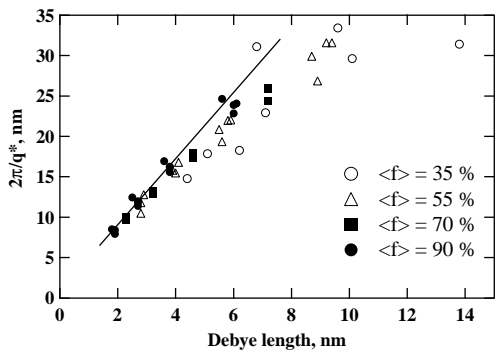}\caption{SAXS correlation length
$2\rm{\pi}/q^{*}$ versus Debye length $\lambda_D$ for various PSS
(see Table) grouped here in four average charge fractions $\langle
f \rangle$. The straight line represents the classical behaviour
in the case of model hydrophilic polyelectrolytes: $2\pi/q^* \sim
\lambda_D$. } \label{Debyelength}
\end{figure}

A series of typical SAXS spectra $I(q)$ are presented in
fig.~\ref{SAXS} for salt-free solutions of PSS as a function of
$f$, all other parameters remaining constant.
One notes the presence of the classical broad maximum also called
the polyelectrolyte peak. In the case of hydrophilic
polyelectrolytes in salt-free water, it is independent of $f$ as
long as $f$ is above the onset of the Manning-Ooasawa condensation
(about 35\% in the case of vinylic
polyelectrolytes)~\cite{Essafi_1999}. Here the situation is very
different. When $f$ decreases, \emph{i.e.} when the hydrophobicity
increases, this broad maximum becomes sharper, more intense and
the maximum moves to smaller $q$ values. In other terms the
characteristic correlation length $2\pi/q^{*}$ increases (in this
example from 11.4 nm to 18.3 nm) with fluctuations decreasing.
Assuming that the correlation length is related to an inter-chain
electrostatic repulsion, its size should be of the order of the
Debye screening length $\lambda_D$. In salt-free solutions
electrostatic screening comes from the residual ions in water (on
the order of $10^{-5} \rm M$) and the free polyelectrolyte
counter-ions at a concentration $C_pf_{eff}$. In our experimental
conditions the former contribution can be neglected and we have:
\begin{equation}
\lambda_{D}\simeq (\frac{\rm{e^2}}{\epsilon {\rm k_B}
T}C_pf_{eff}\rm{N_A})^{-1/2} \label{lambdad}
\end{equation}
where e is the elementary charge, $\rm{\epsilon}$ the dielectric
constant of water, $\rm k_B$ the Boltzman constant, $T$ the
temperature and $\rm N_A$ the Avogadro number. $f_{eff}$ has been
calculated from $f$ assuming that the empirical renormalisation
law established by Essafi \emph{et al.}~\cite{feff_note} is valid
here. $2\pi/q^*$ is plotted in fig.~\ref{Debyelength} as a
function of $\lambda_D$ for all PSS samples.
For clarity of presentation points have been grouped in four
average charge fractions $\langle f \rangle$: 35\%, 55\%, 70\% and
90\%. First of all $2\pi/q^*$ is an increasing function of
$\lambda_D$. This means that the solution is organised on a length
scale of the order of the screening length; therefore it is
strongly related to the free counter-ion concentration. This
confirms indirectly that the anomalously strong reduction of
$f_{eff}$ observed in the dilute regime~\cite{feff_note} also
occurs in this moderately concentrated regimes. However the
proportionality between $2\pi/q^*$ and $\lambda_D$ is only
observed at the highest charge fraction ($\langle f
\rangle=90\%$). In this case PSS behaves as a hydrophilic
polyelectrolyte in the semi-dilute regime and obeys the expected
scaling law: $2\pi/q^*\sim \lambda_D \sim C_p^{1/2}$. Then when
$\langle f \rangle$ decreases and hydrophobicity increases,
$2\pi/q^*$ progressively deviates from the hydrophilic behaviour
($\xi$ proportional to $\lambda_D$, the straight line in
fig.~\ref{Debyelength}). This means that for hydrophobic
polyelectrolytes the solution structure is not solely controlled
by electrostatic interactions. It depends also on the
hydrophobicity of the chain and conformational properties will
have to be taken into account.

In order to ascertain that the observed deviation from the
classical hydrophilic polyelectrolyte behaviour was not due to
some artifacts of the SAXS technique~\cite{SAXS_note}, we have
also measured $\xi$ by AFM. Following the pioneering work of
Milling with fully charged PSS~\cite{Milling_1996}, we have
analyzed the force distance curve of various hydrophobic PSS
solutions confined in the sphere-plane geometry. The polymers were
the same as those used for the SAXS measurements for one chain
length only ($N=410$). Oscillatory forces were systematically
observed and their characterisation will be described in detail in
a forthcoming paper.
\begin{figure}
\onefigure[scale=0.5]{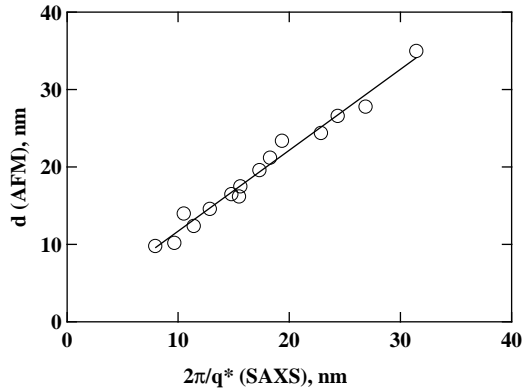}\caption{Direct comparison between
$d$ the period of the oscillating force as measured by AFM  and
the correlation length $2\pi/q^*$ obtained from SAXS. The best fit
line has a slope of 1.05 and an offset of 1.2 nm.} \label{Danvsd}
\end{figure}
We focus here on $d$, the period of oscillation, which has been
shown to be the characteristic length of the solution structure in
real space~\cite{Chatellier_1996,Yethiraj_1999}. The comparison
between $d$ measured by AFM and $2\pi/q^*$ as measured by SAXS is
presented in fig.~\ref{Danvsd}. The two techniques are in
remarkable agreement and within experimental errors we have found:
\begin{equation}
d=1.05\frac{2\pi}{q^*}+1.2 \un{nm} \label{xi_AFMSAXS}
\end{equation}
A detailed comparison between the techniques will be presented
elsewhere.

\begin{figure}
\onefigure[scale=0.5]{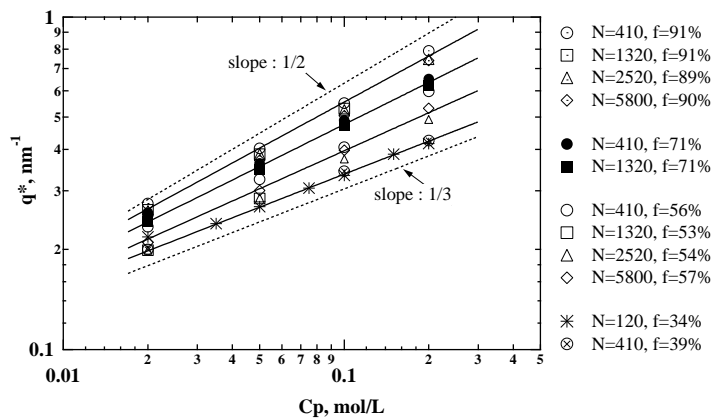} \caption{Position of the correlation
peak q* depending on the polymer concentration $Cp$. Data have
been grouped in four average chemical charge fractions: 35\%,
55\%, 70\% and 90\%. From the top to the bottom straight lines
have slopes of: 0.46, 0.42, 0.38, and 0.33 and the dashed lines
have slopes of 1/2 and 1/3, respectively.} \label{qstar}
\end{figure}

\begin{figure}
\onefigure[scale=0.5]{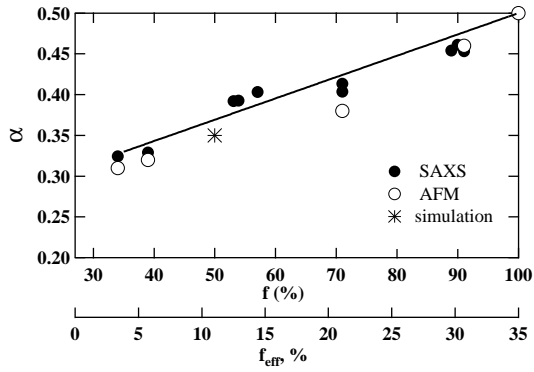} \caption{$\alpha$ exponent in the
power law $\xi\sim C_{p}^{-\alpha}$ depending on the chemical
charge fraction $f$ (upper bottom X-axis) and on the effective
charge fraction $f_{eff}$ (lower bottom X-axis) calculated from
$f$ using~\cite{feff_note}. Filled points are from SAXS, empty
points from AFM and the star is reported from recent
simulation~\cite{Limbach_2002}. The line connects $(f=35\% ;
\alpha=1/3)$ to $(f=100\% ; \alpha=1/2)$. PSS at $f=100\%$ was a
commercial sample purchased from Fluka and dialyzed prior to
measurement.} \label{alpha}
\end{figure}

Within this experimental framework we have then systematically
measured $\xi$ as a function of $N$, $f$, and $C_p$. The classical
logarithmic plot $q^*$ as a function of $C_p$ is presented in
figure~\ref{qstar} for chain lengths ranging from 120 to 5800
monomers and various $f$ per chain length.
First there is clearly no dependence on $N$ as expected for
semi-dilute solutions of overlapping chains. However there is a
strong influence of $f$ even though these polymers are highly
charged strong polyelectrolytes with $f$ above the onset of
Manning-Oosawa condensation ($f\geq35\%$). In the case of
hydrophilic polyelectrolytes we should expect $q^*\sim C_p^{1/2}$
for all $f$. This behaviour is only observed for the highest
charge fraction (at $\langle f \rangle=90\%$ we have $q^*\sim
C_p^{0.46}$) \emph{i.e.} only when PSS almost behaves  as a model
hydrophilic polyelectrolyte. When $f$ decreases, the linear
density of hydrophobic moieties along the chain (styrene monomer)
increases and the intrinsic hydrophobicity of the backbone is
enhanced. The consequence is a progressive deviation towards
$q^*\sim C_p^{1/3}$ at the lowest $f$ ($\langle f \rangle=35\%$).
Having a $C_p$ exponent different from 1/2 is highly unusual for
polyelectrolyte and this clearly results from hydrophobic effects.
In the framework of the pearl-necklace model two limit behaviours
have been predicted ($q^*\sim C_p^{1/2}$ in the string-controlled
regime and $q^*\sim C_p^{1/3}$ in the bead controlled regime) on
each side of a crossover occurring when the pearl size becomes of
the order of $\xi$. In this study such a transition has never been
observed. However it may appear out of the $C_p$ range that is
narrow in this study~\cite{Conc_note}. To sum up, in this regime
where no transition has been observed, a general law for the
position of the peak has been found:
\begin{equation}
q^*\sim N^0C_p^{\alpha}
\end{equation}
This leads for the correlation length $\xi$:
\begin{equation}
\xi\sim N^0C_p^{-\alpha}
\end{equation}
$\alpha$ is plotted in fig.~\ref{alpha} as a fuction of $f$ or
$f_{eff}$ as calculated from $f$ using~\cite{feff_note}.
It has been determined by SAXS and using AFM as well (the
oscillation period that is a measure of $\xi$ scales as: $d\sim
C_p^{-\alpha}$). Both techniques show that $\alpha$ depends on $f$
and continuously decreases from $\alpha=0.50$ at $f=100\%$ to
reach $\alpha=0.33$ at $f=35\%$. At $f=100\%$ PSS behaves as a
hydrophilic polyelectrolyte and obeys the classical power law for
hydrophilic polyelectrolytes in the semi-dilute regime $\xi \sim
C_p^{-1/2}$. At the lowest charge fraction ($f=35\%$), PSS is very
hydrophobic and is thus in very poor solvent condition. The chains
are compact, slightly charged ($f_{eff}=3-5\%$) globules that
electrostatically repel each other. In this case we thus have $\xi
\sim C_{p}^{-1/3}$. The two limit exponents 1/2 and 1/3 can be
physically interpreted: a hydrophilic polyelectrolyte behaviour
when fully charged and a self repelling colloidal organisation at
very low $f$. The intermediate situation when $\alpha$ differs
from 1/3 or 1/2 is very delicate to interpret and it has never
been predicted by any theory. However very recently Limbach
\emph{et al.} made molecular-dynamics simulations on strongly
charged polyelectrolytes in poor solvent which is the situation of
PSS in water. At f=50\%, they found that the position of the peak
in the structure factor scaled as: $q^{*}\sim C_{p}^{0.35}$ for
all investigated $C_p$ ~\cite{Limbach_2002}. This value for
$\alpha$ is reported in fig.~\ref{alpha}. Their result is in
quantitative agreement with the continuous transition that we have
experimentally observed.

\section{Conclusion} The correlation length $\xi$ of well defined hydrophobic
polyectrolyte has been determined using two very different
techniques: SAXS and AFM. $\xi$ obeys a general power law: $\xi
\sim N^{0}C_p^{-\alpha}$. Contrary to the case of hydrophilic
polyelectrolytes, $\alpha$ depends strongly on $f$, decreasing
from 1/2 at $f=100\%$ to 1/3 at $f=35\%$ in the vicinity of the
solubility limit.

\acknowledgments We are grateful to T. A. P. Seery for his
efficient participation in the SAXS experiments. I. Dolbnya and W.
Bras are acknowledged for their assistance at ESRF BM26. We also
wish to thank C. Holm, H.-J. Limbach and S. Biggs for fruitful
discussions.


\begin{thebibliography}{0}
\bibitem{SAXS_note}
In a SAXS experiment, I(q) contains the contribution from three
partial structure factors: monomer/monomer, monomer/counter-ion
and counter-ion/counter-ion. $q^*$ may be displaced.
\bibitem{Conc_note}
Here we focus on the $C_p$ range where the various hydrophobic PSS
are comparable: at high $f$ it is difficult to go at very low
$C_p$ ($<0.01 \un{mol/L}$) because of the poor signal-to-noise
ratio; at low $f$ the maximal $C_p$ value is determined by the
solubility limit.
\end{thebibliography}
\end{document}